\documentclass{article}

\usepackage[T1]{fontenc}
\usepackage{amsmath,amssymb,amsfonts}
\usepackage{graphics, subfigure,color} 
\usepackage{latexsym}
\usepackage[dvips]{graphicx}
\usepackage{epsfig}
\usepackage{cite,hyperref}
\usepackage{chngcntr}

\usepackage[strings]{underscore}

\numberwithin{equation}{section}
\counterwithout{figure}{section}

\usepackage[english]{babel}
\newcommand{\be}{\begin{equation}}
\newcommand{\bee}{\begin{equation}}
\newcommand{\ee}{\end{equation}}
\newcommand{\bea}{\begin{eqnarray}}
\newcommand{\eea}{\end{eqnarray}}

\begin{document}

\title{The Tensor  Track, IV}

%\ShortTitle{Tensor  Track IV}

%
%
\author{Vincent Rivasseau\\
Laboratoire de Physique Th\'eorique, CNRS UMR 8627,\\
Universit\'e Paris XI,  F-91405 Orsay Cedex\\
%E-mail: \email{rivass@th.u-psud.fr}
}

%\author{Another Author\\
%        Affiliation\\
%        E-mail: \email{...}}

\maketitle

\begin{abstract}

%\abstract{
This note is a sequel to the previous series ``Tensor Track I-III''. Assuming 
some familiarity with the tensor track approach to quantum gravity, we
provide a brief introduction to the developments of the last two years and to their corresponding bibliography. 
They center around understanding the interface between random matrices and random tensors through the intermediate field representation, 
finding new types of $1/N$ expansions by enhancing sub-leading tensor interactions, 
exploring the renormalization group flows in the tensor theory space, 
and developing the constructive aspects of the theory.
%} 

\end{abstract}

%\FullConference{Proceedings of the Corfu Summer Institute 2015 ''School and Workshops on Elementary Particle Physics and Gravity''\\
%		1-27 September 2015\\
%		Corfu, Greece}

%\classification{11.10.Gh, 04.60.-m}
%\keywords{Renormalization, tensor models,quantum gravity.}

\section{Introduction}

\medskip
The tensor track \cite{Rivasseau:2011hm,Rivasseau:2012yp,Rivasseau:2013uca} appears as a promising framework 
for a simple and natural\footnote{This naturalness is discussed in \cite{Rivasseau:2016zco} 
in terms of a \emph{quantum relativity principle}.} background-independent ultraviolet-consistent completion of general relativity. Indeed 

\begin{itemize}

\item the tensor theory space \cite{Rivasseau:2014ima} contains triangulations for \emph{every}
piecewise linear manifold in \emph{any} dimension, hence seems a good framework
for the \emph{random-geometric} approach to quantum gravity in dimensions higher than two\footnote{At rank/dimension 2 random tensors reduce to random matrices,
which have been used successfully to quantize gravity in two dimensions.},

\item  functional integration in this space provides a sum over \emph{both} topologies and metrics naturally pondered
by a discrete analog of the Einstein-Hilbert action \cite{ambjorn}, 

\item unexpected asymptotic freedom \cite{BenGeloun:2012pu,BenGeloun:2012yk,Rivasseau:2015ova} 
holds for the simplest renormalizable models, allowing 
analytic understanding of their extreme ultraviolet regime,

\item  constructive methods provide a non-perturbative definition at least of the simplest 
super-renormalizable models \cite{Rivasseau:2016rgt},

\item tensor invariant interactions have allowed to successfully renormalize 
group field theories \cite{Freidel:2005qe,Oriti:2006se,Krajewski:2012aw,Carrozza},

\item field theoretic methods are available to compute numerically 
renormalization group flows and effective actions \cite{Carrozza:2016vsq}.

\end{itemize}

The last point is very important. Indeed a main difficulty is to identify space-time as emergent, hence as a condensate phase of the initial theory. This is 
conceptually similar to the difficult problem of deducing hadronic and nuclear physics from quantum chromodynamics (QCD).
A fully analytic solution should not be expected soon, since in physics effective behaviors qualitatively different from the bare ones 
can almost never be computed analytically. 
Even for the long-time behavior of the three body problem in Newtonian gravity or the 
phase transition of the Ising model in three dimensions (not to mention the formation of
molecules and crystals in the real world), computer simulations are required at some stage. 
Therefore it is expected that the investigation of renormalization group flows and phase transitions in the tensor theory space
 will also require numerical as well as analytic tools. 

\medskip
The subject has now matured. There are several reviews addressing particular 
subtopics and even books published \cite{Carrozza} or in preparation.
Hence this note is not intended as a review but rather 
as an introductory guide to the recent results, structured around four brief sections

\begin{itemize}

\item tensor models,  intermediate field methods and $1/N$ expansions,

\item tensor field theories with and without group field theory  projectors,

\item numerical explorations of their renormalization group flow,

\item constructive results.

\end{itemize}

%
%
%\cite{boulatov,laurentgft,Oriti:2011jm,Krajewski:2012aw,Baratin:2010wi,Baratin:2011aa}, which is the 
%second-quantized, field-theoretic version \cite{Oriti:2013aqa} of loop quantum gravity \cite{Rovelli}, non commutative quantum field theory (NCQFT) 
%\cite{Douglas:2001ba,Rivasseau:2007ab}, random matrix models \cite{Mehta,Di Francesco:1993nw} and (causal) dynamical triangulations \cite{ambjorn-book,scratch,Ambjorn:2013apa}. Based on random tensors and their 
%recently discovered 1/N expansion \cite{Gurau:2011kk,Gur3,GurRiv,Gur4,Gurau:2011xp}\footnote{The theory
%deals with {\emph{a priori unsymmetrized}} tensors; it uses a combinatorial device called color
%to label their indices, and is born out of an attempt to simplify and improve group field theory,
%hence its initial name of {\emph{colored}} group field theory \cite{color,Geloun:2009pe}.}, it proposes to quantize gravity as a quantum field theory (QFT) \emph{of}
%space-time, not \emph{on} space-time, and suggests that our physical space-time emerges\footnote{The concept of \emph{emergent space-time}
%has been attracting increasing attention in the recent years among physicists \cite{Seiberg:2006wf,Oriti:2006ar,Oriti:2007qd,Konopka:2008hp,Sindoni:2011ej}. Like quantum mechanics, 
%it implies a challenging revision of our traditional representation of reality.} 
%as a continuum limit of tensor-type theories through a cascade of phase transitions. Let us start 
%by summarizing the motivations to use random tensors to quantize gravity.

\section{Tensor Models}

%\subsection{Critical Points}

Edge-colored $d$-regular graphs are the basic combinatorial objects of random tensor models, as they provide 
at the same time at rank $d$ the observables and the interactions and at rank $d+1$ the Feynman graphs of the theory 
\cite{Gurau:2009tw,Gurau:2011xp,Gurau:2011kk,Bonzom:2012hw}.
For a recent general review on edge-colored graphs and their basic relation to tensor models see \cite{Ryan}.

In the simplest case these graphs are also required to be bipartite and correspond then to $d$-dimensional orientable geometries \cite{Gurau:2010nd}
and to $U(N)^{\otimes d}$ tensor invariant monomials. See \cite{Carrozza:2015adg} for extension
to the non-bipartite case which corresponds to $O(N)^{\otimes d}$ tensor invariants and can include
non-orientable geometries. Multi-orientable models  correspond to still another tensor invariance 
which is specific to three dimensions. They have been systematically studied by A. Tanasa and collaborators.
 The main recent results establish their full $1/N$ expansion \cite{Fusy:2014rba} 
and their single and double scaling limit \cite{Gurau:2015tua}, a recent review being \cite{Tanasa:2015uhr}. 

Edge-colored graphs attracted early interest in a topological context, since
they are dual to colored triangulations of piecewise linear manifolds
 \cite{FG, FGG, Lins}. The emphasis in the topological 
as well as in the quantum gravity community has been on dimension/rank
3 and 4. But while the topological community, chiefly
interested in classifying and encoding such manifolds \cite{BCCGM,CC1,CC2}, 
has focused on reduction moves allowing to find their simplest colored triangulation, 
the quantum gravity community, chiefly interested in summing
triangulations pondered by the Einstein-Hilbert action, has focused on an almost inverse process,
namely finding infinite families of leading triangulations for this action. It happens that the most important 
family of this type, the melonic parallel/series family \cite{Gurau:2011xp}, in fact reduces through simple moves 
to the unique bipartite $d$-regular graph with two vertices corresponding to the simplest triangulation of the trivial spheric topology. 

Nevertheless classifying and summing are subtly related issues, 
and we can expect progress from dialogue between the two communities, even if 
the typical integers of interest to topologists (regular genus, gem complexity \cite{CC2}),
are different from those of interest to the quantum gravity community, 
such as the Gurau degree which governs the standard tensor $1/N$ expansion \cite{Gurau:2010ba,Gurau:2011aq,Gurau:2011xq}. 
The latter indeed include metric properties of the underlying triangulation 
in addition to its topological properties. 

After the natural single and double scaling limit of tensor models 
has been identified  as branched polymers \cite{GS,Dartois:2013sra,Gurau:2013cbh,Bonzom:2014oua},
an important issue is to go beyond this phase towards more realistic continuum limits. 
This implies understanding better the sub-leading effects 
in tensor models beyond the melonic approximation.
A promising road for this is to analyze the phase transition 
and the symmetry breaking of the $U(N)^{\otimes d}$ symmetry that precisely happens at the 
melonic critical point. This study has started with two recent papers \cite{Delepouve:2015nia,Benedetti:2015ara}.

%\subsection{Intermediate Field Representation and New $1/N$ Expansions}

The intermediate field representation, initially introduced in the subject for constructive purposes \cite{Gurau:2013pca} 
appears more and more as an essential tool for a deeper study of random tensor models
and of tensor field theories. Indeed it provides a bridge between tensor models and the much more 
developed theory of random matrices. In particular quartic melonic models at rank $d$ 
can be represented as a system of $d$ independent commuting random matrices 
coupled via a particular determinant \cite{Gurau:2013pca}.
This representation has been used to probe the spectra of these intermediate field matrices 
and to compute the modifications of their density of states compared
to the usual Wigner's law \cite{Nguyen:2014mga}. It has become also possible to import results
from matrix theory such as Givental identities \cite{Dartois:2014hga} or 
Eynard-Orantin's topological recursion \cite{Dartois:2016bpi} into tensor models.
A review of this far-reaching program is available in the PhD thesis of S. Dartois \cite{Dartois:2015fje}.

Models with higher-than-quartic interactions also admit representations in terms of coupled systems of random matrices 
but these representations in general are more complicated, as it typically involves non-commuting matrices of different sizes.  An important result is that such
representations exist for any invariant and are associated to ``stuffed Walsch maps'' \cite{BLR}. 
Note however that such representations are not unique, as they depend
on the choice of a pairing of the vertices of the invariant. They are a starting point for a general study of enhanced $1/N$
expansions. Rank four tensor models with non melonic quartic interactions enhanced can interpolate between
branched polymers and Brownian spheres, including the ``baby universe'' phase 
at the transition point  \cite{Bonzom:2015axa}. Little is known for models with more than quartic interactions
which may exhibit even richer beahvior. See \cite{Bonzom:2016dwy} for a review of this burgeoning subject.

\section{Field Theories}

Models with tensor invariant interactions and a non invariant propagator have been called tensor field theories. 
The renormalizable models studied so far divide into two main categories, those without \cite{BenGeloun:2011rc,Geloun:2013saa} and with 
\cite{Carrozza:2012uv,Carrozza:2013wda,Carrozza} Boulatov-type group field theory projectors,
which we should nickname respectively as TFTs and TGFT's. TGFTs are TFT's with the particular 
additional ``gauge invariance'' implemented by the Boulatov projector\footnote{Recall that Boulatov projectors were introduced to 
implement the constraint of the BF action in three dimensional quantum gravity. Several generalizations have been proposed 
by loop quantum gravity and group field theory experts to take into account the additional simplicity constraints in the four dimensional case.}. 
They can also be considered as an improvement of the usual group field theories, allowing for their (nevertheless non-obvious) \emph{renormalization}  \cite{Carrozza}. 

The initial research phase was characterized by renormalization theorems at all orders, computations of beta functions at one and two loops approximation
and the rough classification of the corresponding models. The current period
centers around a more systematic investigation of their properties and phase structure, 
generalizing many standard field theoretic tools such as the parametric representation \cite{Geloun:2014ema}, renormalization group equations 
of the Polchinski \cite{Krajewski:2015clk} and Wetterich type \cite{Benedetti:2014qsa,Benedetti:2015yaa}, 
Ward identities  combined with Schwinger-Dyson equations  \cite{Samary:2014tja} and Connes-Kreimer algebras \cite{Avohou:2015sia}.

The most typical result of this second period is the solution of the leading melonic sector of renormalizable 
quartic field theories, which has been obtained both in the TFT \cite{Samary:2014tja,Samary:2014oya} 
and in the TGFT case \cite{Lahoche:2015ola}, 
through closed equations which combine Ward identities and  Schwinger-Dyson equations. Remark that 
such results are the direct analogs in the tensor context of the solution of the leading planar sector of the 
Grosse-Wulkenhaar model in the non-commutative field theory or matrix context \cite{Grosse:2009pa,Grosse:2012uv,Grosse:2014lxa}.

Among other noticeable results are several extensions of TFT's and TGFT's which prepare the ground for future studies. In \cite{Geloun:2015lta}, tensor interactions with ``derivative couplings'',
hence not exactly $U(N)^{\otimes d}$ invariant, have been introduced and investigated. They are an important step for the
development of field theoretic models with enhanced sub-leading interactions in the manner of \cite{Bonzom:2015axa}. Also TGFT's were generalized 
 to the case where the field variable lives on a symmetric space \cite{Lahoche:2015tqa}. 
This is again an important step, preparing the inclusion of simplicity constraints such as Plebanski 
constraints in four dimensional TGFTs.

Recent reviews on the subject are \cite{Geloun:2016bhh,KT2016}.

\section{Renormalization Group Flows}

Investigations of the renormalization group flow in the tensor theory space started with the perturbative computation of the 
beta functions at one or two loops for renormalizable models and led to the the discovery of their generic asymptotic freedom \cite{BenGeloun:2012pu,BenGeloun:2012yk,Rivasseau:2015ova}. 
In the case of models with sixth order interactions such as those of \cite{BenGeloun:2011rc,Carrozza:2013wda}, 
the issue is nevertheless complicated by the presence of two different melonic interactions of order six
and of the relevant quartic interaction; a careful investigation of the rank-three $SU(2)$ case reveals that the flow of the theory slightly misses the apparent Gaussian ultraviolet fixed point
in the quadrant where both six-order melonic coupling constants are both positive \cite{Carrozza:2014rba}. This is however not the full domain of stability of the theory, so that further studies are required.

In \cite{Carrozza:2014rya} the $\epsilon$ expansion of Wilson-Fisher was adapted to this rank-three $SU(2)$ TGFT and a non-trivial fixed point was found in dimension $4- \epsilon$,
suggesting that this theory might be asymptotically safe.

The more recent period has seen the development of the  functional renormalization group (FRG)  in the tensor theory space.
This is a method which can extract some information about the renormalization group
trajectory in a region where the coupling constants are not small. It relies on a different logic than perturbation theory.
Instead of using \emph{a few orders of perturbation theory} (even possibly massaged with tools such as Pade-Borel approximants), 
it performs a truncation of the theory space to \emph{a few operators}\footnote{In practice the truncation starts with a local potential approximation of small overall degree in the fields, 
then eventually adds a few quasi-local operators with derivative couplings, searching for a stable pattern of the flow to emerge as more and more operators are included in the truncation.
Although not fully rigorous, this method typically discovers quickly non-trivial fixed points in simple cases such as Feigenbaum iteration of maps in the interval.} 
but in this reduced space it studies the asymptotic behavior of the corresponding finite dimensional dynamical system under the flow of the truncated renormalization group equation, searching in particular for its fixed points. Usually the RG equation used is 
Wetterich equation \cite{wetterich} since it is a closed equation in terms of one particle irreducible functions such as the self energy, hence easier to analyze than Polchinski's equations.

Like all other renormalization tools the FRG method was invented and applied initially to ordinary quantum field theories, and had to be adapted
to the Einsteinian space of diffeo-invariant actions to take its huge gauge invariance into account \cite{reuter}. A further stage is to adapt it 
to the more abstract, background independent tensor theory space \cite{Rivasseau:2014ima} and to its non-local actions. The first step in this respect 
was taken in \cite{Eichhorn:2013isa} in which the FRG with suitable cutoffs was used to probe the renormalization group flow of Grosse-Wulkenhaar 
models which are matrix models and can also be considered as rank 2 tensor models. In \cite{Eichhorn:2014xaa} this approach was further developed to take into account 
multi-critical fixed points corresponding to the coupling of gravity to conformal matter, and the double scaling limit was also investigated.

The FRG was then adapted to the study of proper tensor models with rank greater than two. In the first paper on the subject \cite{Benedetti:2014qsa}, the simplest rank three renormalizable TFT
with  linear kinetic term and truncation up to quartic melonic interactions was studied. For variables in $U(1)^3$ the dynamical system is non-autonomous (as expected
since the fields take values in a compact space). The ultraviolet and infrared regimes required therefore separate studies. 
Asymptotic freedom in the ultraviolet regime is clearly visible, whether in the infrared 
regime the model exhibits an infrared fixed point which seems to be of the Wilson-Fisher type.

The next steps have consisted in extending the method to TGFTs,  both in rank 3 with a linear kinetic term \cite{Geloun:2015qfa} and in rank 6  with a quadratic kinetic term \cite{Benedetti:2015yaa}
and again at the level of  quartic melonic truncations, essentially confirming the same qualitative behavior of asymptotic freedom in the ultraviolet regime with a fixed point in the infrared.
The decompactification limit where the group $U(1)$ is replaced by ${\mathbb R}$ has then been performed explicitly in \cite{Geloun:2016qyb}, confirming again the existence of
a promising transition to a condensed phase in the infrared regime.

A recent review for this expanding subject is \cite{Carrozza:2016vsq}.

\section{Constructive Results}

Constructive field theory \cite{cqft1,cqft2} allows to circumvent the divergence of perturbative quantum field theory by deriving the physically
interesting quantities such as connected correlation functions from convergent expansions applied directly to the functional integral formulation of the theory.
It can be also considered as a clever repacking of (infinite families of) Feynman graphs.
Tensor constructive results up to now rely on a relatively recent technique called the Loop Vertex Expansion  which combines the 
intermediate field representation with a combinatorial forest formula \cite{Rivasseau:2007fr}.
Although introduced to study constructively random matrices and non-commutative field theories, it is in fact even better adapted to the constructive study of tensor models,
as shown in the pioneering paper \cite{Gurau:2013pca} which established the uniform Borel summability of quartic melonic models at large $N$.

In the case of random tensor models the main recent constructive result is the extension of this uniform Borel summability 
to models with arbitrary quartic interactions (no longer necessarily of the melonic type) \cite{Delepouve:2014bma}.
It required the non-trivial use of iterated Cauchy-Schwarz estimates.

The second main development is the extension of the tensor constructive studies to tensor field theories. In that  case the main goal is
to prove Borel summability of the \emph{renormalized series}. A vigorous constructive program has started, leading to proofs of Borel summability 
for several simple models of the super-renormalizable type both without \cite{Gurau:2013oqa,Delepouve:2014hfa} and with
\cite{Lahoche:2015zya,Lahoche:2015yya} group field theory projectors. It is expected that this program should continue until 
construction of just renormalizable asymptotically free quartic tensor models.

For stable (positive) tensor interactions of order higher than quartic, 
only preliminary results have been obtained \cite{Lionni2016}. They suggest that
there should be an intermediate field representation preserving positivity but 
up to now there is no sign that it can lead to a full-fledged loop vertex expansion.

A recent review of the constructive approach to tensor models can be found in \cite{Rivasseau:2016rgt}.

\medskip\noindent
{\bf Acknowledgments}

\medskip
This review has benefitted from fruitful discussions with many people, in particular R. Avohou, J. Ben Geloun, D. Benedetti,
V. Bonzom, S. Carrozza, S. Dartois, B. Duplantier, B. Eynard, H. Grosse, R. Gurau, T. Krajewski, V. Lahoche, L. Lionni, D. Oriti, 
A. Tanasa, F. Vignes-Tourneret and R. Wulkenhaar.
%I am most indebted to J. Ben Geloun, R. Gurau,  D. Oriti and A. Tanasa for reading the draft of this paper and for suggesting improvements.
I  thank the organizers of the Corfu School for continuing their excellent series and for asking me to prepare these brief notes.

\end{document}